\documentclass[fleqn,usenatbib]{mnras}

\usepackage{newtxtext,newtxmath}

\usepackage[T1]{fontenc}
\usepackage{float}
\usepackage{physics}
\usepackage{subfiles}
\usepackage{comment}
\usepackage{booktabs}
\usepackage{graphics}
\usepackage{comment}
\usepackage{graphicx}
\usepackage{hyperref}
\usepackage{xcolor}
\usepackage{lipsum}
\usepackage{tabularx}
\usepackage{graphicx}
\usepackage{orcidlink}
\usepackage{lipsum}
\usepackage{import}
\DeclareRobustCommand{\VAN}[3]{#2}
\let\VANthebibliography\thebibliography
\def\thebibliography{\DeclareRobustCommand{\VAN}[3]{##3}\VANthebibliography}

\usepackage{graphicx}	
\usepackage{amsmath}	
\usepackage{xcolor}

\definecolor{green_cust}{HTML}{00FF00}

\definecolor{blue_cust}{HTML}{0066ff}

\definecolor{gold_cust}{HTML}{FFBF00}

\definecolor{purple_cust}{HTML}{800080}
\newcommand{\purple}[1]{\textcolor{purple_cust}{#1}}

\definecolor{pink_cust}{HTML}{DC267F}

\definecolor{perfect_green}{HTML}{4FBF26}
\newcommand{\red}[1]{\textcolor{crimson}{#1}}
\definecolor{crimson}{HTML}{DC143C}
\newcommand{\TODO}[1]{{ \textbf{ToDo}: }\{\red{#1}\}}

\newcommand{\LEOF}[1]{{ \textbf{Comment from Leo}: }\{\purple{#1}\}}

\newcommand{\gfo}{AT2017gfo}
\newcommand{\grasp}{{\sc grasp}$^0$ }
\newcommand{\darc}{{\sc darc} }
\newcommand{\code}[1]{{\sc #1}}
\newcommand{\myion}[2]{#1 {\sc \MakeLowercase{#2}}}

\title[Data for Te ions for modelling AT2017gfo]{Electron impact excitation of Te {\sc iv} and {\sc v} and Level Resolved R-matrix Photoionization of Te {\sc i} - {\sc iv} with application to modelling of AT2017gfo}

   \author[Leo P. Mulholland  et al.]{ Leo P. Mulholland\orcidlink{0009-0003-2668-5589}$^{1}$\thanks{lmulholland25@qub.ac.uk}, Catherine A. Ramsbottom\orcidlink{0000-0003-1579-8556}$^{1}$,
  Connor P. Ballance\orcidlink{0000-0003-1693-1793}$^{1}$, \newauthor \hphantom{l}Albert Sneppen\orcidlink{0000-0002-5460-6126}$^{2,3}$ and Stuart A. Sim\orcidlink{0000-0002-9774-1192}$^{1}$. \\
          $^{1}$Astrophysics Research Centre, School of Mathematics \& Physics, Queens University Belfast, BT7 1NN, Northern Ireland.\\
                    $^{2}$Niels Bohr Institute, University of Copenhagen, Jagtvej 128, Copenhagen N, Denmark\\
                    $^{3}$Cosmic Dawn Center, Denmark
          }

\date{Accepted XXX. Received YYY; in original form ZZZ}

\pubyear{\the\year{}}

\begin{document}
\label{firstpage}
\pagerange{\pageref{firstpage}--\pageref{lastpage}}
\maketitle


\begin{abstract}

Spectral modelling of kilonovae (KNe) require large amounts of collisional excitation and photoionization atomic data for lowly ionised (neutral, singly and doubly ionised) species of heavy elements. Much of the data currently used is calculated using approximate hydrogenic results or adopts semi-empirical formulae. We present atomic data for ions of tellurium (Te) computed using the well-known $R$-matrix method. Results will also be presented for radiative and thermal collisions of Te {\sc iv} and {\sc v}, for which the required atomic data are also typically limited in the literature. The Multi-Configuration-Dirac-Hartree-Fock (MCDHF) method is used to produce model atomic structures and radiative rates. These model structures are then used to calculate electron-impact-excitation and photoionization cross-sections. The resulting excitation and radiative rates are further used in a collisional radiative model to produce synthetic spectra, which are compared with observations. We also investigate the possibility of Te {\sc iv} contributing to the 1.08 $\mu$m emission feature in the mid-epochs of \gfo ~alongside the established P-Cygni feature of Sr {\sc ii}.
\end{abstract}

\setlength{\tabcolsep}{4pt}


\section{Introduction}

The spectroscopic observations of AT2017gfo \citep{abbott2017gw170817,smartt2017kilonova,pian2017spectroscopic} and later AT2023vfi \citep{levan2023heavy} has sparked a renewed interest in high $Z$ ($ \gtrapprox 30$) atomic species. In particular, complex numerical simulations of such complicated ejecta compositions require a large amount of atomic data. Outside of local thermodynamic equilibrium (LTE), processes such as ionization, recombination and excitation must be explicitly treated. However, given the expensive nature of complex $R$-matrix calculations 
coupled with the scarcity of atomic data, the elements for which $R$-matrix data is calculated must be prioritized. 

A number of spectral identifications have been proposed, with notable overlap with those elements around the peaks of the rapid-neutron-capture, or $r$-process of nucleosynthesis. First peak elements such as Sr, Y and Zr have been proposed \citep{watson2019identification,sneppen2023discovery,gillanders2024modelling} where complementary atomic data has recently been published \citep{mulholland2024_sr_y_ii}. Studies of potential contributions from heavier elements have also been performed \citep{gillanders2021constraints,Hotokezaka2022WSe,hotokezaka2023tellurium}.
In particular, tellurium (Te, $Z = 52$) has been topical in recent publications \citep{hotokezaka2023tellurium,gillanders2024modelling} and has been attributed to the 2.1 $\mu$m emission feature in the late stages of AT2017gfo. Te lies at the second peak of the $r$-process nucleosynthesis abundance pattern and is thus expected to provide a large proportion of the ejecta mass \citep{hotokezaka2023tellurium,levan2023heavy}.

 In large scale astrophysical modelling, typically semi-empirical formulae have been used to generate large amounts of atomic data for modelling purposes.  For excitation processes, the formulae of \cite{van1962rate} and \cite{Axelrod1980} have been employed \citep{shingles2020,PognanNLTE}. Photoionization data calculated with a hydrogenic approximation \cite{Rybicki1979} has been employed by \cite{pognan22opacities,PognanNLTE} to perform large-scale spectral synthesis using the {\sc sumo} code. It is the goal of this work to expand the growing atomic data sets, to ensure that more accurate and representative spectral synthesis calculations can be carried out with calibrated $R$-matrix data. This method \citep{burke1971electron,burke1975r,burke2011r} has been used extensively in the past to calculate thermal excitation and photoionization data \citep[e.g. ][]{gorczyca1997photoionization,smyth2019level,smyth2019towards,Mulholland24Te}.

We have recently studied the electron-impact excitation of the first three ion stages of Te \citep{Mulholland24Te} and provided the appropriate atomic data that are most important for the modelling of the late optically thin stages of the KNe. Motivated by the presence of relatively high charge states in some simulations due to non-thermal ionization (see e.g \citealt{pognan2022validity}), in this paper we calculate Einstein A-values and electron-impact-excitation rates for Te {\sc iv} and {\sc v}. This will aid modelling across all stages of the KNe, with particular importance in the nebular phase where these are the dominant collisional-radiative processes. Using these structure models for Te {\sc i}-{\sc v}, we additionally provide $R$-matrix level-resolved photoionization cross sections for Te {\sc i}-{\sc iv} for transitions among the ground and metastable levels. Data of this kind is  important in the early stage modelling of the KNe, where the photoionization process is dominant. The full collection of data presented here is expected to be useful in time-dependent modelling, as we have provided data necessary for simulating two different phases of the ejecta dynamics. This paper marks the first $R$-matrix photoionization calculation specifically catered for KNe application, and additionally the first such calculation for the ions of Te.  

 The remainder of this paper is structured as follows. In Section \ref{sec:atomic_structure}, the Dirac-Hartree-Fock methods used to obtain the atomic structure models are reviewed, with the models for Te {\sc iv} and {\sc v} presented. Where possible we have compared with experiment or other calculations for the energy levels and radiative transition data. In Section \ref{sec:excitation} the electron-impact-excitation calculation for Te {\sc iv} and {\sc v} are presented. The relevant collision theory is summarised and some representative collision strengths are presented. These collision strengths have been thermally averaged using a Maxwellian distribution, although the raw collision strengths may be made publicly available to modellers who prefer non-Maxwellian distributions. In Section \ref{sec:crm} collisional radiative modelling is performed using the newly calculated excitation/de-excitation rates. The mid phases of the KNe are discussed, particularly those that feature the emission dominated 1.08 $\mu$m feature which shows remarkable alignment with a ground-state transition of Te {\sc iv}. In Section \ref{sec:photoionization}, the level-resolved Dirac photoionization calculations are discussed, with the theory briefly introduced. The photoionization cross sections for Te {\sc i}- {\sc iv} are presented across a wide range of photon energies, 0-6 Ry. Finally, in Section \ref{sec:conclusion} we conclude and provide an outlook for future work.
\section{Atomic Structure}

\label{sec:atomic_structure}

The \grasp  structure package generates optimised single-electron orbtials based on a set of input configuration-state-functions (CSFs) through the Multi-Configurational-Dirac-Hartree-Fock approach \citep{Grant80,Dyall1989}. The Dirac-Coulomb Hamiltonian (in atomic units) is given by
\begin{equation}
H_{DC} = \sum_i \bigg(c \boldsymbol{\alpha}\cdot\boldsymbol{p}_i + (\beta -I_4)c^2 -\frac{Z}{r_i}\bigg) + \sum_{i>j}\frac{1}{r_{ij}} \label{eq:ham},
\end{equation}
where $\boldsymbol{\alpha}$ and $\beta$ are the set of four Dirac-matrices, $I_4$ is the 4 $\times $ 4 identity matrix, $c$ is the speed of light, $r_i$ is the radial position of electron $i$, $r_{ij}$ is the inter-electron distance and $Z$ is the nuclear charge. The extended-average-level method is employed, which weights the diagonal elements of \eqref{eq:ham} according to the statistical weight of the corresponding configuration in the self-consistent-field optimization procedure. The eigenvalues of this operator naturally correspond to the energy levels of the atomic system. The wavefunctions $\Psi_j$ are then represented by an expansion in terms of the input configurations. Einstein coefficients $A_{j \to i}$ for spontaneous emission from an initial state $i$ to a final state $j$ are calculated as,
\begin{equation}
    A_{j \to i} =  \frac{16\pi^3}{3h\epsilon_0 g_j \lambda^3}|<\Psi_i|O|\Psi_j>|^2,
\end{equation}
for the electric dipole operator $O$, with emitted photon wavelength $\lambda$ and statistical weight of the upper level $g_j$. Here, $h$ and $\epsilon_0$ are Planck's constant and the permittivity of free space. There are similar contributions to the total A-value from higher order multipole operators (see e.g \citealt{cowan1981theory,drake2007springer}). These computed A-values can be adjusted according to spectroscopic wavelengths where possible via,
\begin{equation}
    A_{\text{shifted}} = \Big(\frac{\lambda_{\text{calc}}}{\lambda_{\text{expt}}}\Big)^3 A_{\text{calc}},
\end{equation}
for electric and magnetic dipole transitions with similar relations for higher order transitions, e.g where the wavelength ratio increases to the fifth power for quadrupole transitions. 

Given the intention for these atomic structure models to be carried through to large $R$-matrix calculations, whose computational complexity grows as the number of included configurations to the third power, it is advantageous to keep the atomic structure models relatively small. The included configurations were therefore carefully selected to produce a model representative of the levels important for modelling with minimum size.  

\begin{table}
    \centering
    \begin{tabular}{c l }
    \hline \\

    Te {\sc iv} Model & 4d$^{10}$ 5s$^2$\{4f,5p,5d,6s,6p,6d,7s\};           \\
     - 21 CSF      & 4d$^{10}$ 5s5p\{4f,5d,6s,6p\};           \\
                       & 4d$^{10}$ 5s$\hphantom{^2}$\{5d6s,5d6p,5p$^2$,5d$^2$,6p$^2$\};         \\
                       & 4d$^{10}$ \{5p$^3$, 5p$^2$5d \};  \\
                       & 4d$^{9}$5s$^2$\{5p$^2$,5p5d\};  \\
                       & 4d$^{8}$5s$^2$5p$^3$.  \\
\\
Te {\sc v} Model &  4d$^{10}$ \{5s$^2$, 5p$^2$,5d$^2$,6s$^2$,6p$^2$,6d$^2$,7s$^2$,7p$^2$\}\\
    - 27 CSF      & 4d$^{10}$ 5s\{5p,5d,6s,6p,6d,7s,7p,4f\};           \\
                  & 4d$^{10}$ 5p\{5d,6s,6p,6d,7s\};\\
                  & 4d$^{10}$ 5d\{6s,6p\}; 4d$^{10}$ 6s6p;  \\
                  & 4d$^{9}$5s$^2$\{5p,5d\};  \\
                  & 4d$^{8}$5s$^2$5p$^2$.  \\

    \hline\\
    \end{tabular}
    \caption{Non relativistic valence configurations included in the \grasp calculations for Te {\sc iv} and Te {\sc v}. All configurations shown have a [Kr] core.}
    \label{tab:csfs}
\end{table}

\subsection{Te {\sc iv}}

\begin{figure} 
\centering
    \includegraphics[width = \linewidth]{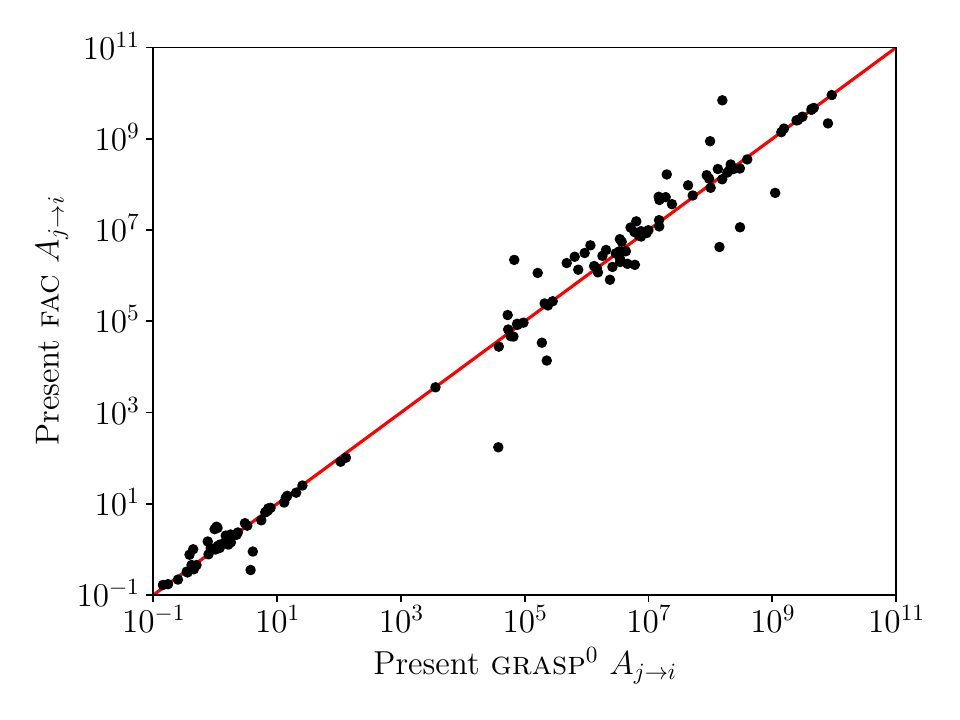}
    \caption{A comparison of {\sc fac} and \grasp Einstein A-values (s$^{-1}$) for Te {\sc iv}.}
    \label{fig:Teiv-avalues}
\end{figure}

In Table \ref{tab:csfs} the 21 non-relativistic configurations included in the wavefunction expansion for Te {\sc iv} are presented. The atomic structure of this charge state has previously been studied by \cite{Crooker:64} with a recent revision by \cite{tauheed1999interpretation}. These two publications interpret two different sets of configurations, which we have compiled here to compare with the present structure model. The comparison of the computed energy levels in Ry relative to the ground state is shown in Table \ref{tab:teiv_energies}. In general, relatively good agreement is found for the lowest 39 energy levels, differences of at best 0.002 and at worst 0.09 Ry are recorded. There is, by contrast, little radiative data in the literature with which to compare, the exception being some forbidden transition probabilities published by \cite{biemont1995forbidden}. A particularly interesting transition is the forbidden decay 5s$^2$5p~$^2$P$^{\rm{o}}_{3/2} \to ~^2$P$^{\rm{o}}_{1/2}$ (Level 2 $\to$ 1,~$\lambda = 1.08~\mu$m). We have calculated $A_{2\to1} = 7.03$ s$^{-1}$ for this line which is in good agreement with the value of $7.09$ s$^{-1}$ published by \cite{biemont1995forbidden}. Clearly the ground term is well represented in the present model, of particular interest in subsequent discussions of the $1.08~\mu$m emission feature in \gfo ~in Section \ref{sec:crm}. Due to the lack of transition data for comparison, the current work completed an additional calculation of A-values using the Flexible Atomic Code ({\sc fac}) of \cite{gu2008flexible}\footnote{{\sc fac} is available from \url{https://github.com/flexible-atomic-code/fac}. Commit \emph{ f19509d} was used in this work.} using the same set of configurations listed in Table \ref{tab:csfs}. This code employs the Dirac-Fock-Slater \citep{Sampson89diracfockslaterref} method and optimizes an analytic radial potential based on the energy of a fictitious mean configuration. The A-values were calculated using energies shifted to spectroscopic wavelengths and compared with those from the present \grasp computations. We compare with this code as it is quite prevalent in the astrophysical community for mass production of data (see e.g \citealt{pognan2022validity,flörs2025calibratedlanthanideatomicdata,dasilva2025systematicbayesianoptimizationatomic}).  The comparison is graphically presented in Fig. \ref{fig:Teiv-avalues}, where good agreement along the line of equality is evident for a range of forbidden and allowed transitions. In particular, excellent agreement is found for some of the weakest lines with transition rates less than 10 s$^{-1}$.
\setlength{\tabcolsep}{2pt}

\begin{table}
    \centering
    \begin{tabular}{lllllrr}
        \toprule
        \hfil Index & \hfil Config. & \hfil Level & \hfil Expt.   & \grasp & $\Delta E$\hphantom{-}  \\ 
        \midrule
        1  & 5s$^2$5p    &$^2$P$^{\rm{o}}_{1/2}$ &  0.0000$^a$ &	0.0000 &          \\
        2  & 5s$^2$5p 	 &$^2$P$^{\rm{o}}_{3/2}$ &  0.0840$^a$ &	0.0780 & -0.0061  \\
        3  & 5s5p$^2$	 &$^4$P$_{1/2}$          &  0.6330$^a$ &	0.6051 & -0.0279  \\
        4  & 5s5p$^2$	 &$^4$P$_{3/2}$          &  0.6722$^a$ &	0.6393 & -0.0330  \\
        5  & 5s5p$^2$	 &$^4$P$_{5/2}$          &  0.7134$^a$ &	0.6799 & -0.0334  \\
        6  & 5s5p$^2$	 &$^2$D$_{3/2}$          &  0.8454$^a$ &	0.8375 & -0.0079  \\
        7  & 5s5p$^2$	 &$^2$D$_{5/2}$          &  0.8640$^a$ &	0.8526 & -0.0114  \\
        8  & 5s5p$^2$	 &$^2$S$_{1/2}$          &  0.9982$^a$ &	1.0650 &  0.0668  \\
        9  & 5s5p$^2$	 &$^2$P$_{1/2}$          &  1.0845$^a$ &	1.1505 &  0.0660  \\
        10 & 5s5p$^2$	 &$^2$P$_{3/2}$          &  1.0931$^a$ &	1.1893 &  0.0962  \\
        11 & 5s$^2$5d	 &$^2$D$_{3/2}$          &  1.1614$^a$ &	1.1540 & -0.0074  \\
        12 & 5s$^2$5d	 &$^2$D$_{5/2}$          &  1.1684$^a$ &	1.1735 &  0.0051  \\
        13 & 5s$^2$6s	 &$^2$S$_{1/2}$          &  1.2162$^a$ &	1.2435 &  0.0273  \\
        14 & 5s$^2$6p    &$^2$P$^{\rm{o}}_{1/2}$ &  1.4703$^b$ &	1.4980 &  0.0278  \\
        15 & 5s$^2$6p    &$^2$P$^{\rm{o}}_{3/2}$ &  1.4941$^b$ &	1.5185 &  0.0244  \\
        16 & 5p$^3$      &$^2$D$^{\rm{o}}_{3/2}$ &  1.6030$^b$ &	1.5663 & -0.0367  \\
        17 & 5s$^2$4f    &$^2$F$^{\rm{o}}_{5/2}$ &  1.6227$^b$ &	1.6273 &  0.0046  \\
        18 & 5s$^2$4f    &$^2$F$^{\rm{o}}_{7/2}$ &  1.6229$^b$ &	1.6256 &  0.0027  \\
        19 & 5p$^3$      &$^2$D$^{\rm{o}}_{5/2}$ &  1.6324$^b$ &	1.5826 & -0.0498  \\
        20 & 5p$^3$      &$^4$S$^{\rm{o}}_{3/2}$ &  1.6444$^b$ &	1.6872 &  0.0427  \\
        21 & 5s5p5d      &$^4$F$^{\rm{o}}_{3/2}$ &  1.7172$^b$ &	1.6404 & -0.0769  \\
        22 & 5s5p5d      &$^4$F$^{\rm{o}}_{5/2}$ &  1.7339$^b$ &	1.6577 & -0.0762  \\
        23 & 5s5p5d      &$^4$F$^{\rm{o}}_{7/2}$ &  1.7598$^b$ &	1.6819 & -0.0779  \\
        24 & 5p$^3$      &$^2$P$^{\rm{o}}_{1/2}$ &  1.7911$^b$ &	1.8246 &  0.0335  \\
        25 & 5s5p5d      &$^4$F$^{\rm{o}}_{9/2}$ &  1.8046$^b$ &	1.7187 & -0.0859  \\
        26 & 5p$^3$      &$^2$P$^{\rm{o}}_{3/2}$ &  1.8059$^b$ &	1.8256 &  0.0197  \\
        27 & 5s5p5d      &$^4$P$^{\rm{o}}_{5/2}$ &  1.8109$^b$ &	1.7618 & -0.0491  \\
        28 & 5s5p5d      &$^4$D$^{\rm{o}}_{1/2}$ &  1.8169$^b$ &	1.7740 & -0.0429  \\
        29 & 5s5p5d      &$^4$D$^{\rm{o}}_{3/2}$ &  1.8206$^b$ &	1.7675 & -0.0531  \\
        30 & 5s$^2$6d	 &$^2$D$_{3/2}$          &  1.8493$^a$ &	1.8715 &  0.0222  \\
        31 & 5s$^2$6d	 &$^2$D$_{5/2}$          &  1.8531$^a$ &	1.8747 &  0.0216  \\
        32 & 5s5p5d      &$^4$D$^{\rm{o}}_{7/2}$ &  1.8656$^b$ &	1.8198 & -0.0458  \\
        33 & 5s5p5d      &$^4$D$^{\rm{o}}_{5/2}$ &  1.8744$^b$ &	1.8248 & -0.0496  \\
        34 & 5s5p6s      &$^4$P$^{\rm{o}}_{1/2}$ &  1.8781$^b$ &	1.8069 & -0.0713  \\
        35 & 5s$^2$7s	 &$^2$S$_{1/2}$          &  1.8785$^a$ &	1.8966 &  0.0181  \\
        36 & 5s5p5d      &$^4$P$^{\rm{o}}_{3/2}$ &  1.8795$^b$ &	1.8210 & -0.0585  \\
        37 & 5s5p5d      &$^2$F$^{\rm{o}}_{5/2}$ &  1.8921$^b$ &	1.8982 &  0.0061  \\
        38 & 5s5p5d      &$^4$P$^{\rm{o}}_{1/2}$ &  1.8945$^b$ &	1.8263 & -0.0682  \\
        39 & 5s5p6s      &$^4$P$^{\rm{o}}_{3/2}$ &  1.9079$^b$ &	1.8389 & -0.0690  \\
    \bottomrule
    \end{tabular}
    \caption{Calculated Te {\sc iv} levels compared with literature experimental values.  Literature values marked with $^a$ are taken from \citet{Crooker:64}, and those marked with $^b$ are from \citet{tauheed1999interpretation}. Energies are presented in Rydberg units (1 Ry $\approx 109,737.316$ cm$^{-1}$)}
    \label{tab:teiv_energies}
\end{table}

\subsection{Te {\sc v}}

By contrast, the spectrum of Te {\sc v} has been studied more recently, both experimentally by \citet{ATauheed_2000}, and theoretically by \cite{ekman2013spectral} via a large scale theoretical atomic structure calculation performed with the {\sc grasp2k} code \citep{jonsson2013new}. The target state energies for the 24 lowest lying levels are presented in Table \ref{tab:tev_energies} and
comparisons are made with the largely experimental work of \cite{ATauheed_2000} and theoretical calculations of \cite{ekman2013spectral}. Note that the levels corresponding to the 5s4f configuration are not experimentally verified and are shown for comparison only.
In general there is good agreement with the experimental energy levels, particularly for the low lying levels which are well represented. For some of the higher lying target states larger energy disagreements of up to 0.07 Ry are found. Given the relatively low temperatures in the observed KNe, it is unlikely that excited Te {\sc V} levels will be populated, but the good representation of the low lying levels gives confidence in any assertions made from modelling KNe or other low-mid temeprature astrophysical plasmas.

\begin{figure} 
\centering
    \includegraphics[width = \linewidth]{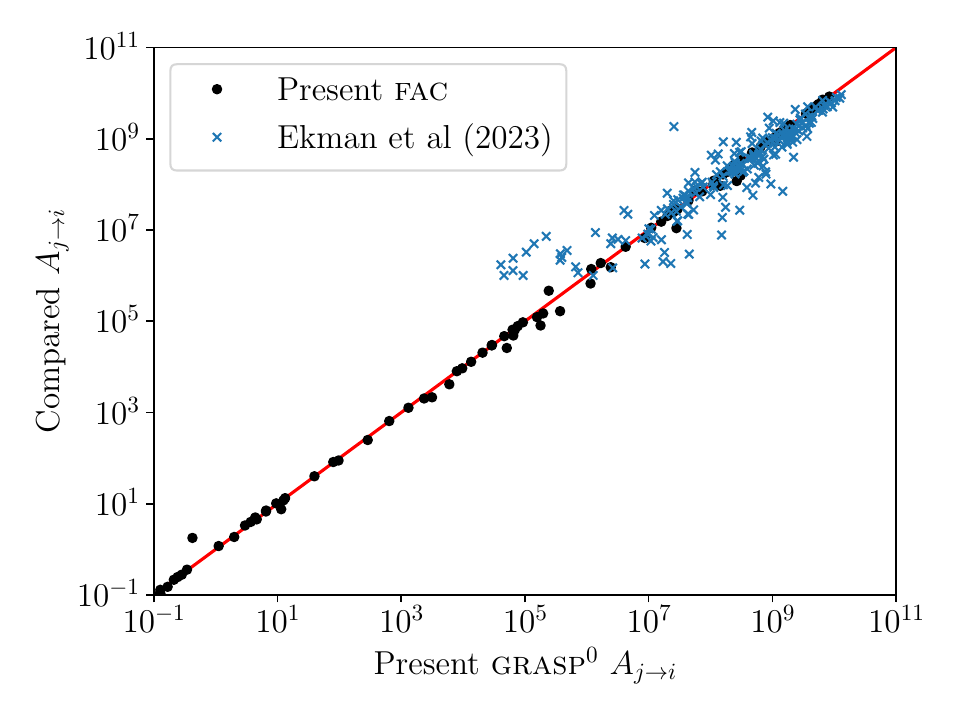}
    \caption{A comparison of A-values (s$^{-1}$) for  Te {\sc V}. The present \grasp calculation (carried through to the {\sc darc} codes) is compared with the theoretical calculations of \citet{ekman2013spectral} as well as values generated in this work with the {\sc fac} code \citep{gu2008flexible}.}
    \label{fig:Tev-avalues}
\end{figure}

The corresponding A-values for transitions among these levels of Te {\sc v} are presented graphically on Fig.  \ref{fig:Tev-avalues}. Again, these transition probabilities have been evaluated using wavelengths which have been calibrated to their experimental positions. Comparisons are made with the {\sc grasp2k} predictions of \citet{ekman2013spectral} and also with an additional calculation completed by the present authors using {\sc fac} and adopting the same set of configurations for Te {\sc v} listed in Table \ref{tab:csfs}, similar to that presented above for Te {\sc iv}. In general all three datasets correlate well along the line of equality, even for the weaker lines presented. This is particularly pleasing given the vastly different sizes of the calculations. \citet{ekman2013spectral} employ a total of around 600,000 $jj$-CSFs across the even and odd symmetries, compared to the compact 267 $jj$-CSFs included in both the present \grasp and {\sc fac} models. The good agreement between the present calculations and a much larger one gives confidence in the subsequent collision computations. It should also be noted that the dataset of \cite{ekman2013spectral} presented on Fig. \ref{fig:Tev-avalues} contains only strong electric-dipole transitions.
\setlength{\tabcolsep}{2pt}

\begin{table}
    \centering
    \begin{tabular}{lllllrrrr}
        \toprule
        \hfil Index & \hfil Config. & \hfil Level & \hfil Expt.   & \grasp & $\Delta E$\hphantom{-}& Ek13 & $\Delta E$\hphantom{-}  \\ 
        \midrule
1	& 5s$^2$	& $^1$S$_0$              &	0.0000	&0.0000 &  & 0.0000 &	  \\
2	& 5s 5p	    & $^3$P$^\mathrm{o}_0$   &	0.6845	&0.6760 &-0.0085 & 0.6854 &	 0.0010 \\
3	& 5s 5p	    & $^3$P$^\mathrm{o}_1$   &	0.7110	&0.7025 &-0.0085 & 0.7115 &	 0.0005 \\
4	& 5s 5p	    & $^3$P$^\mathrm{o}_2$   &	0.7837	&0.7690 &-0.0147 & 0.7833 &	-0.0005 \\
5	& 5s 5p	    & $^1$P$^\mathrm{o}_1$   &	1.0180	&1.0770 & 0.0591 & 1.0112 &	-0.0067 \\
6	& 5p$^2$	& $^3$P$_0$              &	1.6062	&1.6173 & 0.0111 & 1.6030 &	-0.0032 \\
7	& 5p$^2$	& $^3$P$_1$              &	1.6623	&1.6650 & 0.0027 & 1.6577 &	-0.0046 \\
8	& 5p$^2$	& $^1$D$_2$              &	1.6658	&1.6381 &-0.0277 & 1.6653 &	-0.0006 \\
9	& 5p$^2$	& $^3$P$_2$              &	1.7551	&1.7395 &-0.0156 & 1.7517 &	-0.0034 \\
10	& 5p$^2$	& $^1$S$_0$              &	1.9598	&1.9985 & 0.0387 & 1.9619 &	 0.0021 \\
11	& 5s 5d	    & $^3$D$_1$              &	1.9648	&1.9299 &-0.0349 & 1.9647 &	-0.0001 \\
12	& 5s 5d	    & $^3$D$_2$              &	1.9696	&1.9348 &-0.0348 & 1.9667 &	-0.0029 \\
13	& 5s 5d	    & $^3$D$_3$              &	1.9774	&1.9423 &-0.0351 & 1.9753 &	-0.0021 \\
14	& 5s 5d	    & $^1$D$_2$              &	2.1146	&2.1662 & 0.0515 & 2.1138 &	-0.0008 \\
15	& 5s 6s	    & $^3$S$_1$              &	2.1947	&2.1277 &-0.0670 & 2.1951 &	 0.0004 \\
16	& 5s 6s	    & $^1$S$_0$              &	2.2486	&2.1937 &-0.0548 & 2.2507 &	 0.0021 \\
17	& 5s 4f	    & $^3$F$^{\mathrm{o}}_2$ &	2.2953*	&2.3705 & 0.0752 & 2.3602 &	 0.0649 \\
18	& 5s 4f	    & $^3$F$^{\mathrm{o}}_3$ &	2.2966*	&2.3726 & 0.0760 & 2.3603 &	 0.0637 \\
19	& 5s 4f	    & $^3$F$^{\mathrm{o}}_4$ &	2.2983*	&2.3757 & 0.0775 & 2.3608 &	 0.0626 \\
20	& 5s 4f	    & $^1$F$^{\mathrm{o}}_3$ &	2.3286*	&2.3992 & 0.0706 & 2.3907 &	 0.0620 \\
21	& 5s 6p	    & $^3$P$^{\mathrm{o}}_0$ &	2.4968	&2.4259 &-0.0709 & 2.4966 &	-0.0003 \\
22	& 5s 6p	    & $^3$P$^{\mathrm{o}}_1$ &	2.5006	&2.4314 &-0.0692 & 2.5002 &	-0.0004 \\
23	& 5s 6p	    & $^3$P$^{\mathrm{o}}_2$ &	2.5288	&2.4567 &-0.0722 & 2.5280 &	-0.0008 \\
24	& 5s 6p	    & $^1$P$^{\mathrm{o}}_1$ &	2.5397	&2.4757 &-0.0640 & 2.5389 &	-0.0008 \\
    \bottomrule
    \end{tabular}
    \caption{The first 24 calculated Te {\sc v} levels compared with the  values of \citet{ATauheed_2000}, all of which except the 5s4f (marked *) were determined via experiment. Also shown are the levels calculated by \citet{ekman2013spectral}  (Ek13) with the corresponding differences to the \citet{ATauheed_2000} levels. Energies are presented in Rydberg units (1 Ry $\approx 109,737.316$ cm$^{-1}$).}
    \label{tab:tev_energies}
\end{table}

\section{Electron-Impact Excitation} \label{sec:excitation}

\begin{figure*} 
\centering
    \includegraphics[width = 0.85\linewidth]{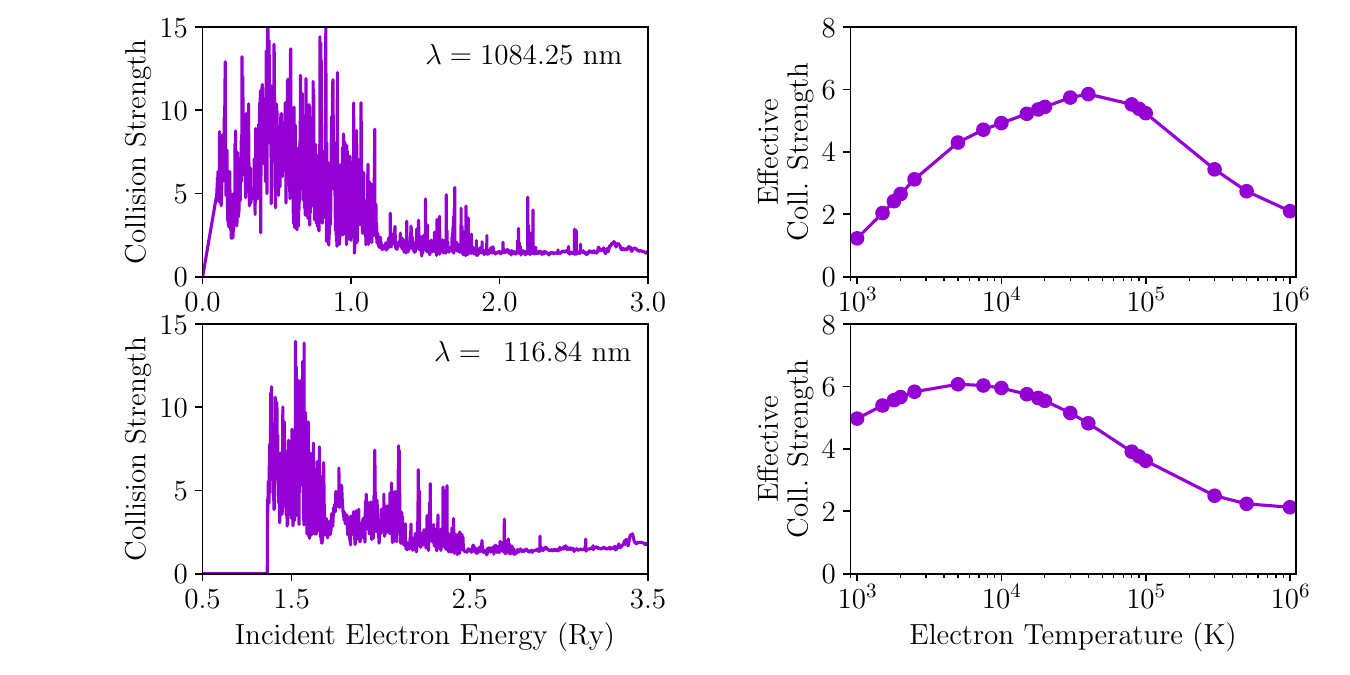}
    \caption{The collision strengths as a function of incident electron energy for two Te {\sc iv} transitions. The two top panels show the raw and Maxwellian averaged collision strengths for the 5s$^2$5p    $^2$P$^{\rm{o}}_{1/2}$ $\to$ $^2$P$^{\rm{o}}_{3/2}$ ($\lambda = 1084.25$nm) forbidden transition. The bottom two panels show the raw and Maxwellian averaged collision strengths for the 5s$^2$5p    $^2$P$^{\rm{o}}_{3/2}$ $\to$ 5s5p$^2$ $^2$D$_{5/2}$ ($\lambda = 116.84$nm) intercombination line. }
    \label{fig:Te3+:exc}
\end{figure*}
 
\begin{figure*} 
\centering
    \includegraphics[width = 0.85\linewidth]{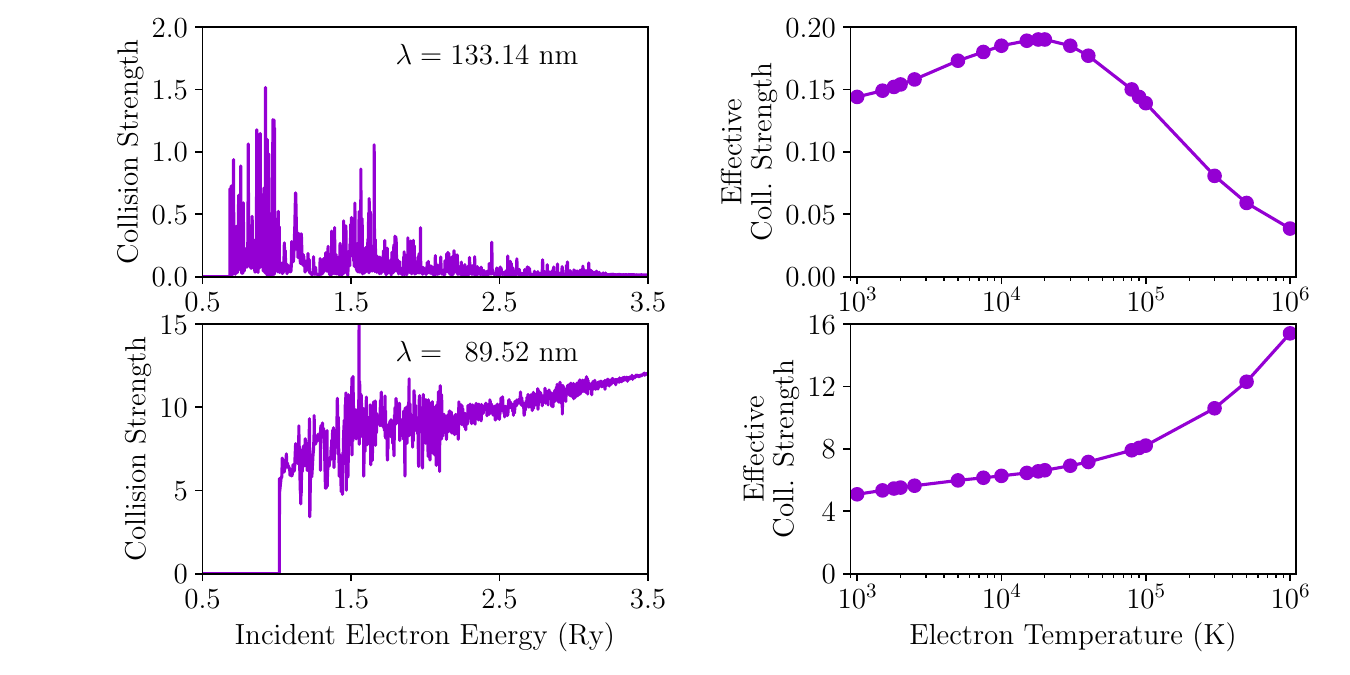}
    \caption{The collision strengths as a function of incident electron energy for two Te {\sc v} transitions. The two top panels show the raw and Maxwellian averaged collision strengths for the 5s$^2$    $^1$S$_{0}$ $\to$ 5s5p $^3$P$^{\rm{o}}_{0}$ ($\lambda = 133.14$nm) intercombination  transition. The bottom two panels show the raw and Maxwellian averaged collision strengths for the 5s$^2$    $^1$S$_{0}$ $\to$ 5s5p $^1$P$^{\rm{o}}_{1}$ ($\lambda = 89.52$nm) allowed line.}
    \label{fig:Te4+:exc}
\end{figure*}

Electron-impact-excitation rates are important for any modelling of astrophysical plasmas that do not assume the approximation of local thermal equilibrium. Excitation/de-excitation rates for Te {\sc i} - {\sc iii} have been published recently by \citet{Mulholland24Te} for use in NLTE modelling of KNe. Here we present the corresponding calculations for Te {\sc iv} and {\sc v} for completeness. The pertinent details surrounding the $R$-matrix theory of these calculations may be found in \citet{burke2011r} and will not be reproduced here. 

A collision strength for each transition $\Omega_{ij}$ is computed as a sum over contributions from all included partial waves. This is a measure of the strength of the transition as a function of incident electron energy. Contributions from high $J$ partial waves are approximated by a top-up rule described in \citep{burgess1974coulomb}. For use in astrophysical applications these collision strengths are typically averaged according to a Maxwellian thermal distribution, 
\begin{equation}
    \Upsilon_{ij}(T_e) = \int_0^{\infty} \Omega_{ij}(\epsilon_j)e^{-\epsilon_j / kT_e}\hspace{1mm} \text{d}\left(\epsilon_j / kT_e\right) \label{eq:effective_cs},
\end{equation}
where $T_e$ is the electron temperature of the plasma and $\epsilon_j$ is the post-collision energy of the incoming electron. The calculation of these integrals is supplemented by the use of large-energy limits of the $\Omega_{ij}$ for dipole transitions which have monotonically increasing behaviour at large $\epsilon_j$. These effective collision strengths can subsequently be converted to excitation and de-excitation rates,
\begin{align*}
       q_{i\to j}(T_e) &= \frac{8.63\times10^{-6}}{g_i T_e^{1/2}} \Upsilon_{ij}(T_e) e^{-E_{ij}/kT_e} \hspace{1mm}\label{eq:rates}, \\
    q_{j \to i}(T_e) &= \frac{g_i}{g_j} e^{E_{ij} /kT_e} q_{i\to j},
\end{align*}
which can be readily be employed in a collisional-radiative model (see for example \citealt{Summers2006}). The important details of the collision calculations for each ion considered are summarised below. Note that in both calculations the target levels are shifted to the experimental energies presented in Tables \ref{tab:teiv_energies} and \ref{tab:tev_energies} (except for the 5s4f levels of Te {\sc v} for which no experimental levels are available, the calculated \grasp levels were retained in this case). In the remainder of this section we show collision strengths for a selection of transitions, although transitions between all included levels may be made available to the reader.

\subsection{Te {\sc iv}}

The close-coupling wavefunction expansion for the Te {\sc iv} excitation calculation employed 20 continuum basis orbitals and an $R-$matrix boundary of $a = 18.08$ a.u. With 100 target levels retained in the expansion, a maximum size of the $N+1$ Hamiltonian matrix of $11075 \times 11075$ was obtained. The first 32 partial waves with $2J = 0-30$ (both even and odd parity) were calculated with 8,000 total incident energy points. The energy spacing was $4.5 \times 10^{-4}$ Ry covering a range of 0-3.6 Ry. The remaining partial waves $2J=32-62$ (with top-up) were computed on a grid of 800 points with a mesh size of $4.5 \times 10^{-3}$ Ry. On Fig. \ref{fig:Te3+:exc}, the collision strengths and Maxwellian integrated effective collision strengths for the forbidden 5s$^2$5p $^2$P$^{\rm{o}}_{1/2}$ $\to$ $^2$P$^{\rm{o}}_{3/2}$ and allowed 5s$^2$5p    $^2$P$^{\rm{o}}_{3/2}$ $\to$ 5s5p$^2$ $^2$D$_{5/2}$ transitions are shown as a function of incident electron energy and temperature respectively.

\subsection{Te {\sc v}}

For the Te {\sc v} excitation calculation we also included 20 continuum basis orbitals, and an $R-$matrix boundary was set at $a = 17.94$ a.u. The expansion retained 100 target levels resulting in a maximum size of the $N+1$ Hamiltonian matrix of $9307 \times 9307$. The low partial waves with $2J=1 - 31$ were calculated with a fine mesh of 8,000 points and an energy spacing of $8 \times 10^{-4}$ Ry to adequately capture the resonance structure of the collision profile. These energy parameters cover a large energy range of 0-6.4 Ry. By contrast for the high partial waves with $2J=33 - 63$, 800 points and an energy spacing of $8 \times 10^{-3}$ Ry was adopted. The large energy range covered combined with the calculated asymptotic behaviour for the dipole transitions, as well as the large number of partial waves gives confidence that the collision strengths and thus the integrals \eqref{eq:effective_cs} have converged.

On Fig. \ref{fig:Te4+:exc} we present the 5s$^2$\hphantom{1}$^1$S$_0$ $\to$ 5s5p\hphantom{1}$^3$P$^{\rm{o}}_0$ intercombination transition and the 5s$^2$\hphantom{1}$^1$S$_0$ $\to$ 5s5p\hphantom{1}$^1$P$^{\rm{o}}_1$ allowed transition. The collision strengths as a function of incident electron energy in Ry and the corresponding effective collision strengths as a function of electron temperature in Kelvin are again shown.

\section{Collisional-Radiative-Modelling} \label{sec:crm}

The calculated electron-impact excitation rates were employed in the collisional radiative modelling {\sc ColRadPy} package \cite{johnson2019colradpy}. In principle such models can also include photoionization cross-sections such as those presented in Section \ref{sec:photoionization}, however this is outside the aims of the current work and is reserved for a more detailed and sophisticated radiative transfer study in the future. We follow the same analysis procedure as detailed in \cite{mulholland2024_sr_y_ii,Mulholland24Te}.

Of particular interest in this work is the Te {\sc iv} 5s$^2$5p~$^2$P$^{\rm{o}}_{3/2} \to ^2$P$^{\rm{o}}_{1/2}$ forbidden transition. It is noteworthy this transition has been identified previously in the context of planetary nebulae \citep{sterling2020neutron}. The wavelength of this line is $1.08~\mu$m and coincides with a feature in \gfo, which becomes emission dominated at late times ($\gtrsim 5$ days) and appears to fade away by $\sim$ 8 days. This feature has been investigated previously in terms of a group of Sr {\sc ii} allowed transitions. The early time P-Cygni feature at this wavelength is agreed to be likely caused by these Sr {\sc ii} lines \citep{watson2019identification,sneppen2024emergence,sneppen2024rapid}. 
It was found that the Sr {\sc ii} transition could faithfully reproduce the emission feature at around 7 days, but also exhibited additional emission features towards the blue end of the spectrum that are not present in the spectra of \gfo ~\citep{gillanders2024modelling}. It was suggested further by \cite{gillanders2024modelling} that microclumping of the ejecta is one possible avenue to remove the forbidden emission features. It could also be explained by overlying opacity. It is also possible that the feature itself is fading from an earlier time through a delay caused by reverberation (McNeill et al. submitted to MNRAS). In the event these effects cannot explain the differences between simple models and the real spectra, it is possible that another transition is blending in and potentially becoming important at later times. There has also been discussion in the literature of a He interpretation of the feature \citep{tarumi2023nonlteanalysisheliumstrontium,sneppen2024heliumfeaturesinconsistentspectral,sneppen2024heliumindicatorneutronstarmerger}. 

\begin{figure}
    \centering
    \includegraphics[width =\linewidth]{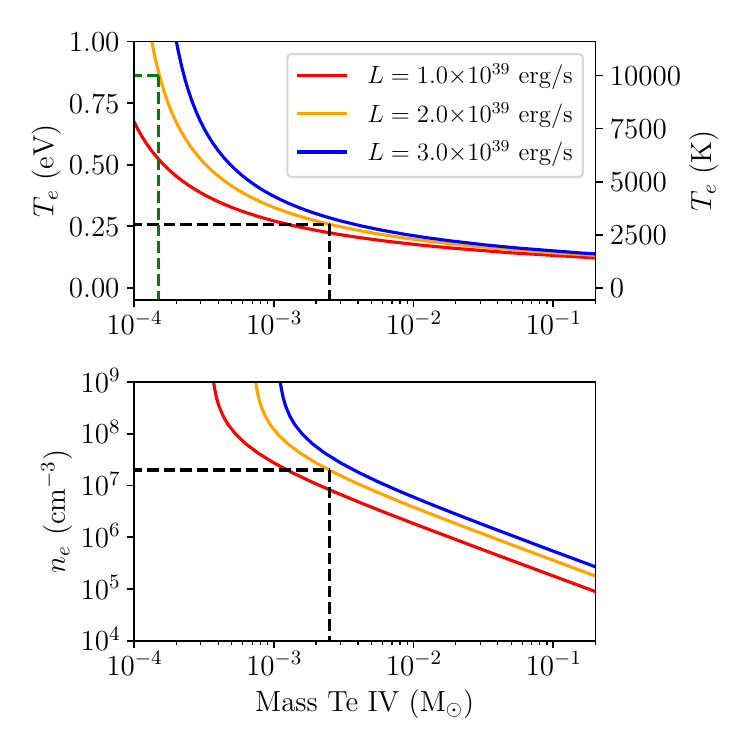}
    \caption{Contour plots of constant luminosity, at values of $L = 1.0\times 10^{39}$ erg/s, $L = 2.0\times 10^{39}$ erg/s, $L = 3.0\times 10^{39}$ erg/s. The parameter space of Te {\sc iv}, electron temperature and electron density are explored. On the top panel, the density is fixed at 2$\times 10^{7}$ cm$^{-3}$ and temperature and mass are varied. The dashed lines are fixed electron temperatures, namely 3,000K (0.26 eV, black) and 10,000K (0.86 eV, green). 
    On the bottom panel, the electron temperature is fixed at 3,000K and density and mass are varied. The black dashed line is a  density of 2$\times 10^{7}$ cm$^{-3}$. On both figures the black dashed lines correspond to a mass estimate of $2.5 \times 10^{-3}$M$_{\odot}$. On the top figure, the green dashed line corresponds to a mass of  $1.5 \times 10^{-4}$M$_{\odot}$.}
    \label{fig:te3+_contour}
\end{figure}

In this section we investigate the possibility of Te {\sc iv} contributing to the emission dominated phases of the feature.  At 7.4 days post explosion, \cite{gillanders2024modelling} integrate this feature and estimate a line luminosity of  $\sim$ 2.0 $\times 10^{39}$ erg s$^{-1}$. At a temperature of $3000$K and a density of $2.0 \times 10^{7}$ cm$^{-3}$, we find that the Te {\sc iv} mass required to reproduce the feature was $\sim 2.5 \times 10^{-3}$ M$_{\odot}$.  While perhaps high, this mass estimate lies at a feasible order of magnitude. A large Te mass is motivated as this element lies at the second peak of the $r$-process. The parameter space of temperature, density and mass requirement is explored in Fig. \ref{fig:te3+_contour}, where contours of constant line luminosity are shown. This places the Te {\sc iv} mass in a similar regime to that of Te {\sc iii} that has been calculated previously by \cite{hotokezaka2023tellurium} and \cite{Mulholland24Te}.

\begin{figure}
    \centering
    \includegraphics[width =\linewidth]{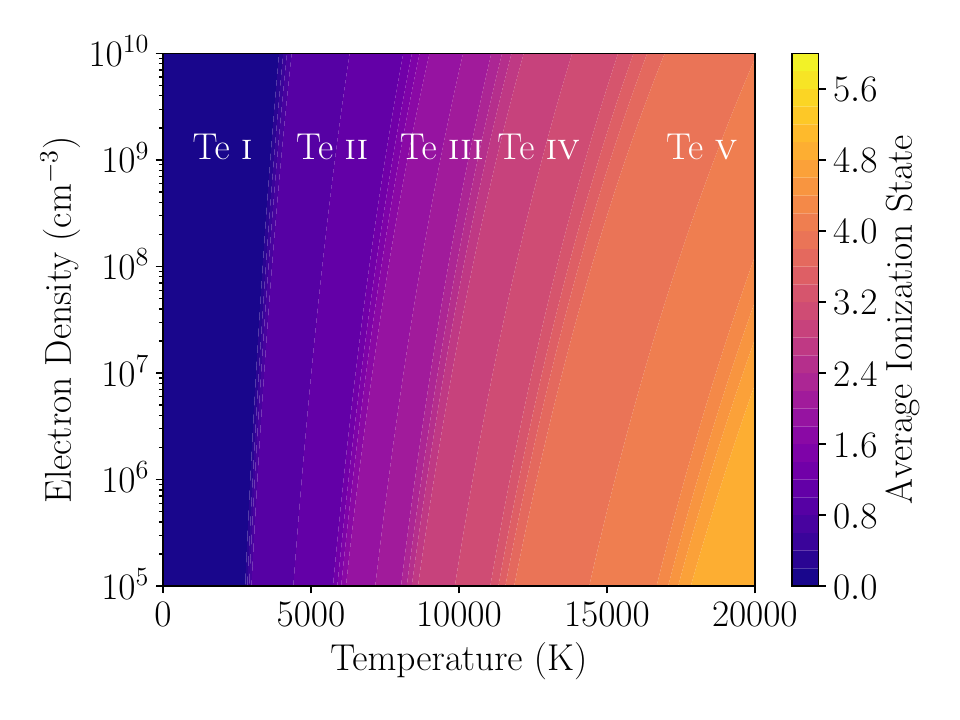}
    \caption{Average degree of ionization as a function of electron density and temperature. The Saha-Boltzmann law of LTE ionization is employed.}
    \label{fig:saha}
\end{figure}

A caveat of this analysis is the feasibility of the presence of a triply ionized species at this relatively low temperature. In LTE with only thermal ionization, the Saha-Boltzmann law requires a characteristic temperature of around 8,500K for the amount of Te {\sc iv} to be significant when compared to that of Te {\sc iii} at this density. A heatmap of mean ion stage as a function of density and temperature using the Saha-Boltzmann law is shown on Fig. \ref{fig:saha}. These temperatures are considerably hotter than those previously reported at this epoch for \gfo ~\citep{hotokezaka2023tellurium,sneppen2024emergence}. Incidentally, at temperatures this high - Te {\sc iv} will not exhibit any contaminant lines at other wavelengths, although Te {\sc iii} may do so.

LTE is, however, unlikely to provide an accurate ionization estimate at late times. Collisions with non-thermal electrons due to the thermalisation of radioactive-decay particles is expected to be a significant contribution to the ionization balance.  This is exemplified by simulations such as those of \cite{pognan2022validity}, where large amounts of Te {\sc iv} begin to appear around 7-10 days. This is consistent with results for Neodymium presented in \cite{brethauer2025nonthermalionizationkilonovaejecta}, where Nd {\sc iv} appears in the outer ejecta around the same time frame. It is therefore possible that around this epoch we can potentially observe emission from higher charge states. This said, stronger conclusions are reserved for radiative transfer simulations employing calibrated collision data, such as those presented here, along with detailed ionization balance including non-thermal electrons.

Additionally, it is seen that Te {\sc iii} also exhibits a line at around 1.08 $\mu$m, corresponding to the 5p$^2$ $^1$D$_2$ $\to$ 5p$^2$ $^3$P$_2$ forbidden transition. This is evident at relatively high temperatures  but was not found to contribute to the feature without also grossly overestimating the 2.1 $\mu$m feature at masses consistent with literature values. However, it is possible with high temperatures and low ion masses that this could produce emission consistent with \gfo.

\begin{figure}
    \centering
    \includegraphics[width =\linewidth]{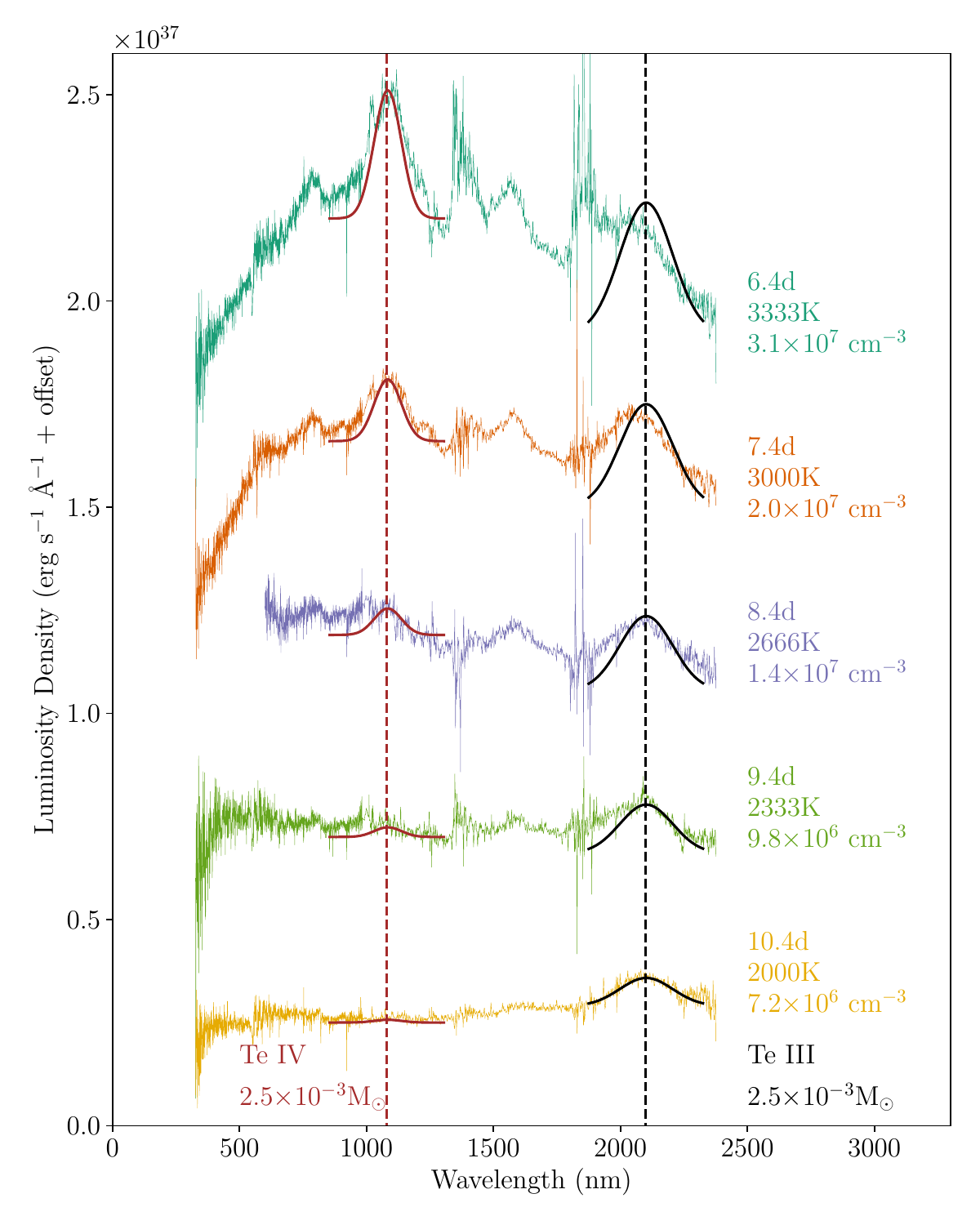}
    \caption{Synthetic spectra of Te {\sc iii} and Te {\sc iv} compared with the spectra of \gfo. The ion mass is fixed at 2.5 $\times 10^{-3}$ M$_\odot$. The electron temperature is taken to be a linear decrease, with the density following a $t^{-3}$ decay with time.}
    \label{fig:manydays}
\end{figure}

To investigate the consistency of this identification with the spectral evolution of \gfo, we show broadened spectra of Te {\sc iii} and {\sc iv} for  6.4d - 10.4d with $\beta = 0.07c$ (which is the same value used for Te {\sc iii} in \citealt{Mulholland24Te} and results in a FWHM = 0.126 $\mu$m for 1.08 $\mu$m),  compared to observation is also shown on Fig. \ref{fig:manydays}. 
For simplicity, the ionization state is assumed to be constant - and the electron density is thus assumed to follow a $t^{-3}$ evolution. With few contaminant lines from these two ions, the mass and temperature are degenerate - i.e varying the mass and temperature give similar effects. For this reason we elect to keep the ion mass constant and vary the temperature. In reality, both likely vary differently than is shown on Fig. \ref{fig:manydays}. The temperature of 3,000K at 7.4d is chosen to give the best 
agreement with the spectra. The remaining temperatures assume a linear cooling to show under this fairly extreme condition that the spectra is consistent with both Te {\sc iii} and {\sc iv} emission. In general these temperatures are close to blackbody inferences made in works such as \cite{hotokezaka2023tellurium}. Invoking much higher temperatures for the single element produces rich emission structures not seen in the observations. This would hint towards a relatively cool electron-temperature, although we emphasise that there is of course large freedom in assumed abundances and relative ionisation states.

It can be seen that the central wavelength matches that of the feature very well.   Given a consistent temperature and density profile that explains the 2.1 $\mu$m feature with Te {\sc iii}, adopting this for Te {\sc iv} shows a feature at 1.08 $\mu$m at 6.4-8.4d, but not at subsequent epochs. It is then possible in the intermediate time frame of around 7 days, that Te {\sc iv} contributes to the 1.08 $\mu$m emission before fading out - with the ejecta retaining the mass of Te {\sc iii} required to produce the 2.1 $\mu$m emission feature. A time series of spectra of \gfo~ are compared with synthetic spectra of Te {\sc iv} and {\sc iii} at 6.4d - 10.4d days post explosion on Fig. \ref{fig:manydays}. For simplicity of comparison, the mass of Te {\sc iv} is taken to be constant for the which is perhaps valid as the ion fraction is unlikely to change considerably in a relatively short time scale. Again, such conclusions are reserved for more detailed calculations including a robust ionization balance. It is evident that at the temperatures where Te {\sc iii} can feasibly produce the 2.1 $\mu$m feature, a comparable amount of Te {\sc iv} produces no feature that stands out from the noise in the later epochs. In this sense, the notion of the 1.08 $\mu$m feature containing contribution from Te {\sc iv} is feasible, as its contribution is somewhat consistent with the fading of the feature. 



To summarise, the Te {\sc iv} identification presents an advantage over one of pure Sr {\sc ii}, namely that at temperatures as high as 10,000K, Te {\sc iv} will still only exhibit this 1.08 $\mu$m emission feature with no pollution from other emission lines, however Te {\sc iii} will have pollutant emission lines. While the line is producible by this ion at relatively low temperatures, its existence at this temperature seems unlikely, as typically ionization processes also heat the plasma. Regardless, should Te {\sc iv} be present in the plasma, it is quite plausible that it could be a significant contributor to the emission dominated phases that feature in \gfo, while ions with allowed lines around this wavelength such as Sr {\sc ii} providing the P-Cygni behaviour in the earlier phase. More conclusive quantitative statements can perhaps be made in the future with expanding atomic data sets including recombination rates and more detailed evolution models beyond the scope of this work. Of additional future interest is the further exploration of non-thermal ionization effects with future simulations, which may inflate the average degree of ionization in observationally important elements.

\section{Photoionization}\label{sec:photoionization}

\begin{figure*} 
\centering
    \includegraphics[width = \linewidth]{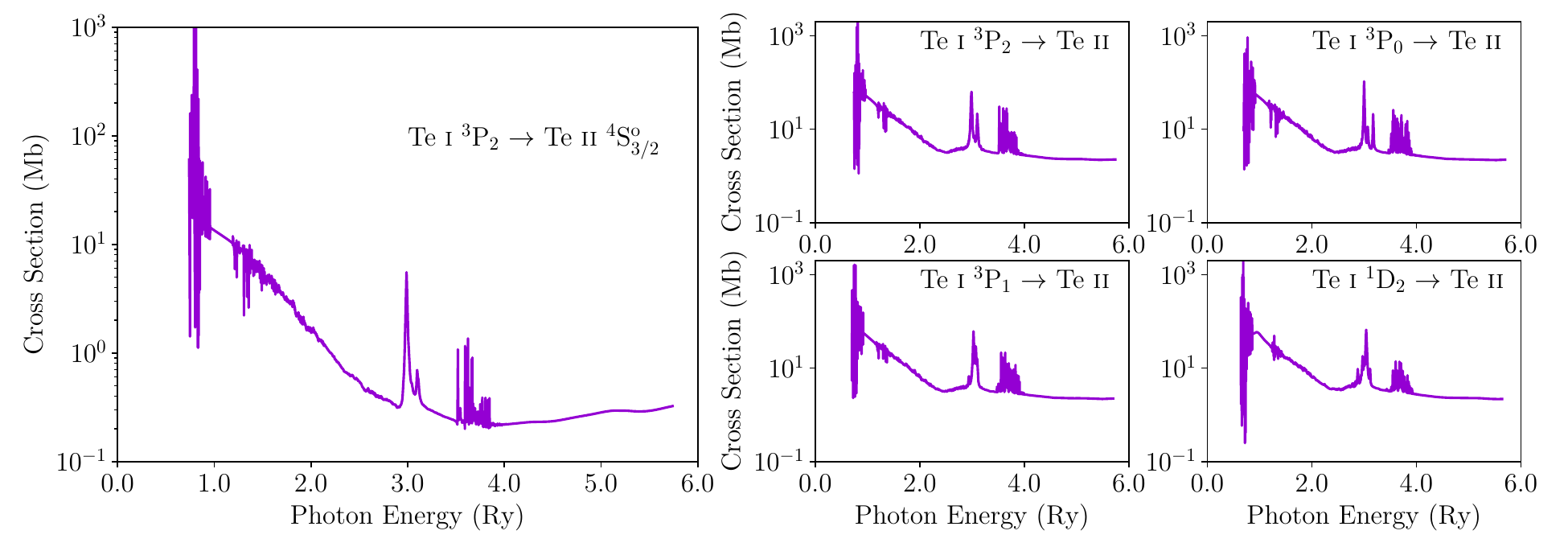}
    \caption{Photoionization cross sections for \myion{Te}{i}. The left hand panel shows the cross section for the ground state of \myion{Te}{i} being ionized to the ground state of \myion{Te}{ii}. The four right hand panel shows the ionization for the ground and first three fine structure levels, where the cross sections are summed over all \myion{Te}{ii} levels included in the calculation. }
    \label{fig:Te0+:pi_ground}
\end{figure*}

\begin{figure*} 
\centering
    \includegraphics[width = \linewidth]{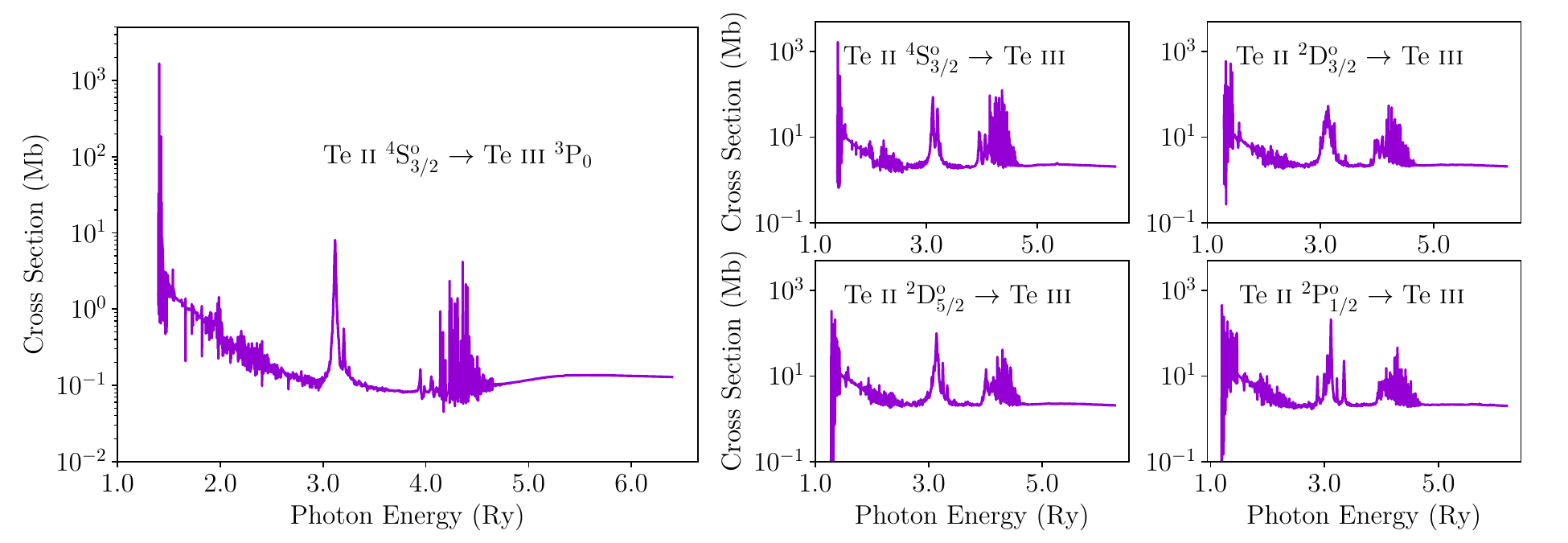}
    \caption{Photoionization cross sections for \myion{Te}{ii}. The left hand panel shows the cross section for the ground state of \myion{Te}{ii} being ionized to the ground state of \myion{Te}{iii}. The four right hand panel shows the ionization for the ground and first three fine structure levels, where the cross sections are summed over all \myion{Te}{iii} levels included in the calculation. }
    \label{fig:Te1+:pi_ground}
\end{figure*}

\begin{figure*} 
\centering
    \includegraphics[width = \linewidth]{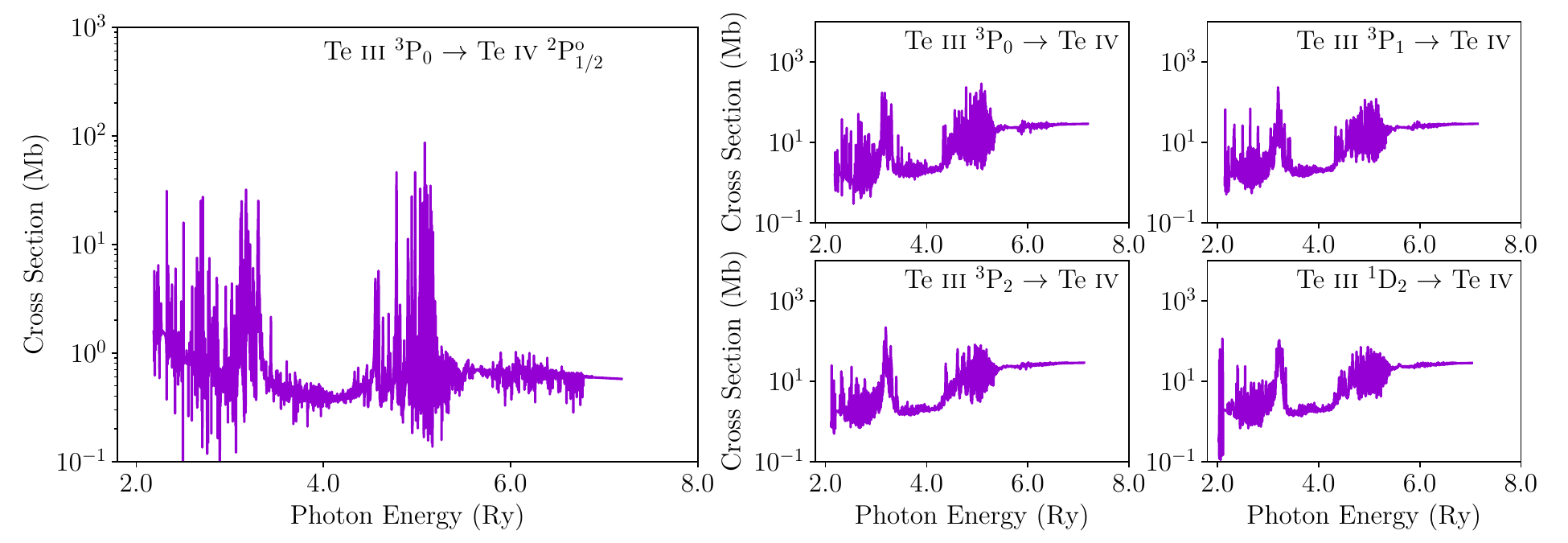}
    \caption{Photoionization cross sections for \myion{Te}{iii}. The left hand panel shows the cross section for the ground state of \myion{Te}{iii} being ionized to the ground state of \myion{Te}{iv}. The four right hand panel shows the ionization for the ground and first three fine structure levels, where the cross sections are summed over all \myion{Te}{iv} levels included in the calculation. }
    \label{fig:Te2+:pi_ground}
\end{figure*}

\begin{figure*} 
\centering
    \includegraphics[width = \linewidth]{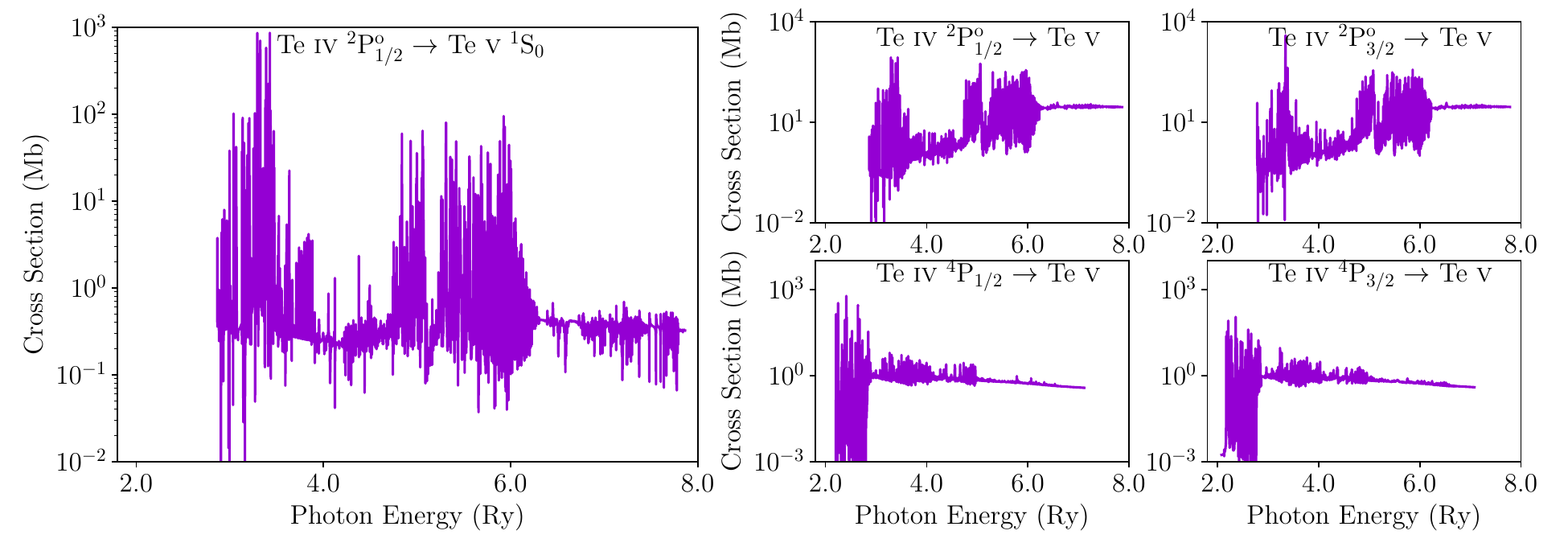}
    \caption{Photoionization cross sections for \myion{Te}{iv}. The left hand panel shows the cross section for the ground state of \myion{Te}{iv} being ionized to the ground state of \myion{Te}{v}. The four right hand panel shows the ionization for the ground and first three fine structure levels, where the cross sections are summed over all \myion{Te}{v} levels included in the calculation.}
    \label{fig:Te3+:pi_ground}
\end{figure*}

Photoionization is a neighbouring process to thermal excitation, and has been postulated to be important in the early stages of the Kilonovae where the plasma is thought to be radiation dominated \citep{kasen2017origin}. Additionally, photoionization has been found to be very important for other astrophysical events such as supernovae \citep{shingles2022modelling}. More generally, spectral synthesis codes such as {\sc cloudy} \citep{gunasekera202323} require the input of large amounts of photoionization data.

In addition to the excitation calculations we also present total and level resolved photoionization cross sections, making use of the Te {\sc iv} and {\sc v} target models discussed previously and the Te {\sc i}, {\sc ii} and {\sc iii} structures presented in \citep{Mulholland24Te}. The {\sc darc} \citep{Ballance2020_DARC} implementation uses optimized orbitals and configuration interaction for the $N$-electron target to calculate photoionization of the $N+1$-electron parent ion using the $R$-matrix method. The associated theory can be found in \cite{burke2011r} and will not be reproduced here. In general, the photoionization cross section is given by matrix elements of the dipole operator $D_g$,

\begin{equation}
    \sigma_{\text{PI}} = A \frac{8\pi^2 \alpha a_0^2 }{3g_i} \Big| \bra{\Psi^{J\pi,-}_f}D_g\ket{\Psi^{J\pi}_i} \Big|^2,
\end{equation}
where $\ket{\Psi^{J\pi}_i}$ is the state of the $N+1$ bound system, and $\ket{\Psi^{J\pi,-}_f}$ is the continuum state of the $N$-electron residual ion and ejected electron. In the length guage, the factor $A = \omega$, and in the velocity gauge $A = 1/\omega$, where $\omega$ is the incident energy of the photon in Ry. Here, $g_i$ is the statistical weight of the initial bound state,  and $a_0$ is the Bohr radius. A target for the $N$-electron (residual ion) system is constructed via the methods described in the previous section, and the $N+1$ bound system is described using a close-coupling expansion. By constructing and diagonalising the $N+1$ Hamiltonian, the bound states of the parent system can be approximately obtained for each requested $J\pi$ and the transition elements with the dipole operator $D_g$ can be calculated.   $R$-matrix calculations such as these allow for the full treatment of resonances which are not present in other more approximate methods. We therefore expect the photoionization cross sections described here to allow for an accurate description of the relevant opacities, as well as for use in large scale modelling codes. In the remainder of this section we show cross sections for a selection of transitions calculated, although all transitions between included levels will be made available to the reader.

\subsection{Te {\sc i}}

Using the Te {\sc ii} target described in \cite{Mulholland24Te}, a model photoionization of Te {\sc i} was constructed using dipole transitions up to an initial state of 2$J = 8$. A continuum basis size of 22 was employed with the $R$-matrix boundary set at $r = 29.76$ a.u. The level expansion retained 629 Te {\sc ii} energy levels, the first 49 of which were shifted to their spectroscopic values. This resulted in a maximum Hamiltonian matrix size of $69482 \times 69482$. This large number of states were retained in this expansion to ensure convergence of the resonance structure up to around 3 $Ry$.  For the photoionization cross sections, a mesh grid of $3.90\times 10^{-4}$ Ry was used to span the incident photon energy range. The parent Te {\sc i} bound-states constructed resulted in a calculated ground-state ionization energy of $0.739$ Ry, which is a slight overestimate compared to the recommended value of $0.662$ Ry \citep{nist}.

The photoionisation cross section of ground state Te {\sc i} to ground state Te {\sc ii} is shown on the left panel of Fig. \ref{fig:Te0+:pi_ground}. The photoionization of the ground and first three metastables, summed over all Te {\sc ii} states included in the calculation are shown on the four right panels. Note that the cross-sections exhibits a broad resonance around 3.05 Ry, which appears to correspond to a photo-excitation-of-core transition of the 5s inner shell. 

\subsection{Te {\sc ii}}

The target structure of Te {\sc iii} described in \cite{Mulholland24Te} was employed to calculate photoionization cross sections of Te {\sc ii}, with the addition of the 4d$^9$5s$^2$5p$^3$ and 4d$^{10}$5s5p$^2$6p configurations included to capture more inner-shell ionizations of the Te {\sc ii} parent. Dipole transitions were included up to an initial state 2$J = 11$ with a continuum basis size of 22 per angular momentum. The $R$-matrix boundary was set at $19.8$ a.u. 379 Te {\sc iii} levels were retained in the wavefunction expansion, with the first 30 shifted to spectroscopic values (except for the 5p5d $^3$F$^{\rm{o}}_4$ level, see \citealt{Mulholland24Te}). This gave rise to Hamiltonian matrices of a maximum size of $56298\times 56298$.  The ionisation energy of Te {\sc ii} was predicted to be $1.40$ Ry, in good agreement with the NIST recommended value of $1.37$ Ry \citep{nist}. In the outer region, a fine mesh of 12800 points was employed with an energy spacing of $3.90 \times 10^{-4}$ Ry. 

The photoionisation cross section of ground state Te {\sc ii} to ground state Te {\sc iii} is shown on the left panel of Fig. \ref{fig:Te1+:pi_ground}. The photoionization of the ground and first three metastables, summed over all Te {\sc iii} states included in the calculation are shown on the four right panels.
Note that the cross-sections exhibits a broad resonance around 3.0 Ry, which appears to correspond to a photo-excitation-of-core transition of the 5s inner shell.

\subsection{Te {\sc iii}}

The target structure of Te {\sc iv} described in Section \ref{sec:atomic_structure} was employed to calculate photoionization cross sections of Te {\sc iii}. Dipole transitions were included up to an initial state 2$J = 8$ with a continuum basis size of 30 per angular momentum. This gave rise to Hamiltonian matrices of a maximum size of $14785\times14785$. The $R$-matrix boundary was set at $18.1$ a.u. Of the 445 Te {\sc iv} levels calculated, 100 were retained in the wavefunction expansion. As described before, the first 39 were shifted to their spectroscopic values. The ionisation energy of Te {\sc iii} was calculated as $2.099$ Ry. This shows reasonable agreement with the NIST recommended value of $2.046$ Ry \citep{nist}. In the outer region, a fine mesh of 12800 points was employed with an energy spacing of $3.90 \times 10^{-4}$ Ry. 

The photoionisation cross section of ground state Te {\sc iii} to ground state Te {\sc iv} is shown on the left panel of Fig. \ref{fig:Te2+:pi_ground}. The photoionization of the ground and first three excited levels, summed over all Te {\sc iv} states included in the calculation are shown on the four right panels.

\subsection{Te {\sc iv}}

Finally, the target structure of Te {\sc v} described in Section \ref{sec:atomic_structure} was employed to calculate photoionization cross sections of Te {\sc iv}. For dipole transitions up to an initial state of 2$J = 9$, the maximum Hamiltonian size was $9307 \times 9307$. The $R$-matrix boundary was set at $17.9$ a.u., and 100 target levels were retained in the wavefunction close-coupling expansion. The photoionisation cross section of ground state Te {\sc iv} to ground state Te {\sc v} is shown on the left panel of Fig. \ref{fig:Te3+:pi_ground}. The photoionization of the ground and first three excited level, summed over all Te {\sc v} states included in the calculation are shown on the four right panels. The target levels of Te {\sc v} were shifted to their spectroscopic values (except the 5s 4f levels as discussed previously).

\section{Conclusions and Outlook}

In this paper, the atomic structures of Te {\sc iv} and {\sc v} were calculated using the \grasp code. Energy levels and transition rates were calculated. These were employed in the {\sc darc} codes to calculate the corresponding electron-impact-excitation rates. These will be made available at \cite{openadas_site} and in the supplementary material. 

It has been pointed out that the first forbidden line of Te {\sc iv} is coincident with the 1.0~$\mu$m emission feature in the mid epochs of \gfo. It was suggested that while Sr {\sc ii} remains the mostly likely candidate for the earlier P-Cygni feature at this wavelength \citep{watson2019identification,gillanders2024modelling}, it is possible that the later strong emission feature is contributed to by Te {\sc iv}. The existence of such a triply ionized species at this early time has been discussed, and we have concluded that it is feasible under very certain conditions of the ejecta evolution, within the simple model presented here. The modelling of \cite{pognan2022validity} shows some circumstances in which the existence of the ion is potentially possible. It was found that under NLTE conditions Te {\sc iv} is able to produce the line luminosity presented by \cite{gillanders2024modelling} at very reasonable points in the parameter space of electron temperature, density and ion mass. Additionally this identification remains valid at relatively high temperatures as there will be no other significant emission from Te {\sc iv}. The only major caveat of this identification is the plausibility of the ionization at these times, which may be studied further in future works with detailed non-thermal ionization balance. By contrast, little evidence of observable signatures of Te {\sc v} are found, but the data has been presented nonetheless for completeness.

The atomic structures presented in this work and in \cite{Mulholland24Te} have been employed in the {\sc darc} codes to produce $R$-matrix level resolved photoionization cross sections in a fully $jj$-coupled setting. To our knowledge this calculation marks the first of its kind for Te ions. It is intended the photoionization and thermal exictation data presented here will be used in large scale modelling codes by the KNe community. Additionally this allows for the potential phasing out of approximate hydrogenic photoionization cross sections typically used in such codes at present.


\label{sec:conclusion}

\section*{Funding}

Funded/Co-funded by the European Union (ERC, HEAVYMETAL, 101071865). Views and opinions expressed are however those of the author(s) only and do not necessarily reflect those of the European Union or the European Research Council. Neither the European Union nor the granting authority can be held responsible for them.

\section*{Acknowledgements}\label{sec:acknowledgements}

We thank our colleagues at Queen's University Belfast and Auburn University for helpful discussion. We are grateful for use of the computing resources from the Northern Ireland High Performance Computing (NI-HPC) service funded by EPSRC (EP/T022175). 

This work used the DiRAC Data Intensive service (DIaL3) at the University of Leicester, managed by the University of Leicester Research Computing Service on behalf of the STFC DiRAC HPC Facility (www.dirac.ac.uk). The DiRAC service at Leicester was funded by BEIS, UKRI and STFC capital funding and STFC operations grants. DiRAC is part of the UKRI Digital Research Infrastructure.

This work used the DiRAC Extreme Scaling service (Tursa) at the University of Edinburgh, managed by the EPCC on behalf of the STFC DiRAC HPC Facility (www.dirac.ac.uk). The DiRAC service at Edinburgh was funded by BEIS, UKRI and STFC capital funding and STFC operations grants. DiRAC is part of the UKRI Digital Research Infrastructure.

This research used resources of the National Energy Research Scientific Computing Center(NERSC), a Department of Energy Office of Science User Facility using NERSC award FESERCAP0031596.

\section*{Data Availability}

The effective collision strengths for Te IV and V are recorded in the standard adf04 format on both this articles online supplementary material and at http://open.adas.ac.uk/. Photoionization cross sections will be available in the supplementary data. Other data underlying this article will be shared on reasonable request to the corresponding author.



\bibliographystyle{mnras}
\bibliography{bibliography} 





\bsp	
\label{lastpage}
\end{document}